\documentstyle[aps,preprint,epsfig]{revtex}
\begin{document}
\title
{Double parton scatterings in b-quark pairs production at the
LHC}
\author{A. Del Fabbro and D.Treleani}
\address{ Dipartimento di Fisica Teorica dell'Universit\`a di Trieste and
INFN, Sezione di Trieste,\\ Strada Costiera 11, Miramare-Grignano,
I-34014 Trieste, Italy.} \maketitle
\begin{abstract}
A sizable rate of events where two pairs of $b$-quarks are
produced contemporarily is foreseen at the CERN LHC, as a
consequence of the large parton luminosity. At very high energies
both single and the double parton scatterings contribute to the
process, the latter mechanisms, although power suppressed, giving
the dominant contribution to the integrated cross section.
\end{abstract}
\vspace{3cm}
E-mail delfabbr@trieste.infn.it \\
E-mail daniel@trieste.infn.it \\
\newpage
\section{Introduction}
One of the main topics at the LHC is the production of $b$-quarks,
both to search for CP violation, looking at $b$ decays, and to
test QCD, by studying the production
mechanism\cite{Catani:2000jh}. Bottom quarks are also a large
source of background to several processes of interest, as in
various promising channels for Higgs
detection\cite{Kunszt:1996yp}. The production mechanism of heavy
quarks in hadronic collisions pose, on the other hand, non trivial
problems already at smaller energies, where also the simplest
observable quantity, the integrated inclusive cross section, is
not reproduced trivially.

The inclusive cross section of $b$-quarks production has been
evaluated in pQCD at the next to leading order in
$\alpha_S$\cite{Nason:1987xz}. Unfortunately comparisons with the
recent experimental data of the D0
Collaboration\cite{Abbott:1999se} at TEVATRON have shown that the
NLO pQCD calculations underestimate the cross section by a factor
$\sim 2,3$, showing that NNLO corrections, whose explicit
evaluation is still an open question, give a large contribution to
the cross section.

\noindent A complementary approach to heavy quarks production,
which keeps explicitly into account that transverse momenta and
virtualities of the interacting partons become increasingly
important in the kinematical regime of $s \gg m_b^2\sim \hat s\gg
\Lambda^2$, and includes terms at every order in $\alpha_S$ in the
calculation of the cross section, is the $k_t$-factorization,
where the interaction is factorized into un-integrated structure
functions and off shell matrix elements
~\cite{Catani:1990eg}\cite{Collins:1991ty}\cite{Gribov:tu}. From
the phenomenological point of view $k_t$-factorization is not
inconsistent with HERA and TEVATRON data, allowing one to
reproduce both the value of the integrated inclusive cross section
and various differential distributions, including the correlation
in the azimuthal angle between the produced $b$ quarks, where
different approaches are less successfully compared with
experiment (see \cite{Anderson:2002cf} and
references therein).

Interestingly, although the value of the integrated inclusive
cross section cannot be obtained trivially, one may find several
cases where the overall effect of higher order corrections amounts
to a simple rescaling of the lowest order parton model result.
There are in fact several distributions, derived either using the
$k_t$-factorization approach or by working out the cross section
at the NLO pQCD, which are rather similar (apart form the
normalization factor) to those obtained with a simplest lowest
order calculation\cite{Ryskin:2000bz}. Hence, in a few cases, the
whole effect of higher order corrections is (approximately)
reduced to a single numerical value, the so-called $K$ factor:
\begin{equation}
K\,=\,\frac{\sigma(b{\bar b})}{\sigma_{LO}(b{\bar b})}\,.
\end{equation}
where $\sigma(b{\bar b})$ is the inclusive cross section for
$b{\bar b}$ production and $\sigma_{LO}(b{\bar b})$ the result of
the lowest order calculation in pQCD.

When looking at extrapolations of the cross section at high
energies, one finds that the result is affected by several
uncertainties, as the knowledge of the parton structure functions
at very small $x$ and the values of the heavy quark mass and of
the running coupling constant. Although the expected inclusive
cross section of $b$ production is hence still pretty uncertain at
LHC energy, all estimates point in the direction of rather large
values, as a consequence of the high parton
luminosity\cite{Catani:2000jh}. The fairly large flux of partons
make it also plausible to expect a sizable rate of events, where
two or more $b{\bar b}$ pairs are produced contemporarily by
different partonic collisions in a given $pp$
interaction\cite{dpth}. Although at present stage all quantitative
predictions for this much more structured interaction process are
unavoidably pretty uncertain, the large cross sections foreseen at
the LHC is, in our opinion, a strong motivation to make an attempt
of giving a few quantitative indications on the production rate of
multiple $b{\bar b}$ pairs trough multiparton interactions at the
LHC, comparing with the rates to be expected by the more
conventional single parton scattering mechanism.

Since the details of the elementary production of heavy quarks are
still a matter of debate, we limit our considerations, for the
production of multiple $b{\bar b}$ pairs, to the simplest cases,
where the whole effect of higher order corrections is taken into
account by the overall normalization factor. Given the lack of
information on higher order corrections in the $2\to4$ processes,
we make moreover the assumption that the $K$ factors of the $gg\to
b{\bar b}b{\bar b}$ and of the $gg\to b{\bar b}$ processes are
equal. Hence we work out the $gg\to b{\bar b}$ process in the
$k_t$-factorization approach, fixing the input parameters by
comparing with the TEVATRON data, and extrapolate the cross
section at LHC energies, identifying a few distributions where the
effect of higher order corrections reduceds to a simple rescaling
of the lowest order result. The value of the $K$-factor derived in
this way is then used to renormalize the double $(gg\to b{\bar
b})^2$ and the single $gg\to b{\bar b}b{\bar b}$ parton scattering
cross sections, which we evaluate by working out all Feynman
diagrams at order $\alpha_S^4$.

\section{$ b\bar b$ cross section at TEVATRON and LHC
and $K$-factor}

In the $k_t$-factorization approach the $b{\bar b}$ production
cross section is expressed as\cite{Catani:1990eg}\cite{Collins:1991ty}
\begin{equation}
\sigma(pp\rightarrow b\bar b)\,=\,\int \frac{d^2q_{t1}}{\pi}
\frac{d^2q_{t2}}{\pi}\, dx_1
dx_2\, f(x_1,q_{t1},\mu) f(x_2,q_{t2},\mu)
\,\hat\sigma(x_1,q_{t1};x_2,q_{t2};\mu)\,
\end{equation}
where $f(x,q_t,\mu)$ is the unintegrated structure function,
representing the probability to find a parton  with momentum
fraction $x$, transverse momentum $q_t$ at the factorization scale
$\mu$, while $\hat\sigma$ is the off-shell partonic cross section
of the process $g^*g^*\rightarrow Q\bar Q$.

To work out the inclusive cross section we use two different
prescriptions for constructing the $k_t$-distributions from the
usual integrated parton densities. The first prescription is based
on the conventional DGLAP evolutions
equations\cite{Kimber:1999xc}, with virtual corrections re-summed
in the survival probability factor $T_a(k_t^2, \mu^2)$
\cite{Marchesini:1987cf}. Hence the un-integrated structure
function for the parton $a$ reads
\begin{equation}
f_a (x, k_t^2, \mu^2) \; = \; T_a (k_t^2, \mu^2) \left [
\frac{\alpha_S (k_t^2)}{2 \pi} \: \int_x^{1 - \delta} \:
P_{aa^\prime} (z) \: a^\prime \left ( \frac{x}{z}, k_t^2 \right )
dz \right ]\,
\end{equation}
where $P_{aa^\prime} (z)$ is the splitting function, $a^\prime
\left ( x, k_t^2 \right )$ the integrated structure function and
$\delta$ a cutoff parameter introduced to give sense to the
integral. Although not written explicitly also $T_a (k_t^2,
\mu^2)$ depends on $\delta$, in such a way the $f_a (x, k_t^2,
\mu^2)$ is a smooth function when $\delta$ becomes small.

As for the second prescription we follow ref.\cite{blumlein},
where the un-integrated structure functions are obtained from the
leading order BFKL equation and are expressed as the convolution
of the usual collinear gluon densities $G(x,\mu^2)$ with the
universal function ${\mathcal G}(x,k_t^2,\mu^2)$
\begin{equation}
{\mathcal F}(x,k_t^2,\mu^2)\,=\,\int^1_x\, d\xi\, {\mathcal
G}(\xi,k_t^2,\mu^2) \,G({\xi\over x},\mu^2)
\end{equation}
The weight factors ${\mathcal G}(\xi,k_t^2,\mu^2)$ have
 a known analytic expression, in
double-logarithmic approximation, in terms of Bessel functions and
depend on the quantity $\bar \alpha_s=3\alpha_s/\pi$, which in the
BFKL formalism is a fixed parameter, related to the pomeron
intercept $\alpha(0)=1+\Delta$, where $\Delta$ in leading log
approximation is $\Delta=4 \bar \alpha_s\log 2$.
Following\cite{Baranov:ja} we take the value $\Delta=0.35$.

To generate the  un-integrated structure functions
 we have used the parton distributions set GRV94\cite{Gluck:1994uf}
  with  factorization scale $\mu_F^2=\hat s$.
Hence in  evaluating the cross sections, in the
$k_t$-factorization approach,
 we have set the renormalization scale equal to the gluon virtuality.
To obtain the cross section at the lowest order in pQCD we have
used the MRS99 parton distributions~\cite{mrs99}, with
factorization and renormalization scale equal to the transverse
mass of the $b$-quark. Comparing  the total cross section values
in the two approaches, we have obtained for the $K$-factor the
value $K\sim 5.5$.

\begin{figure}[t]
\begin{center}
\epsfig{figure=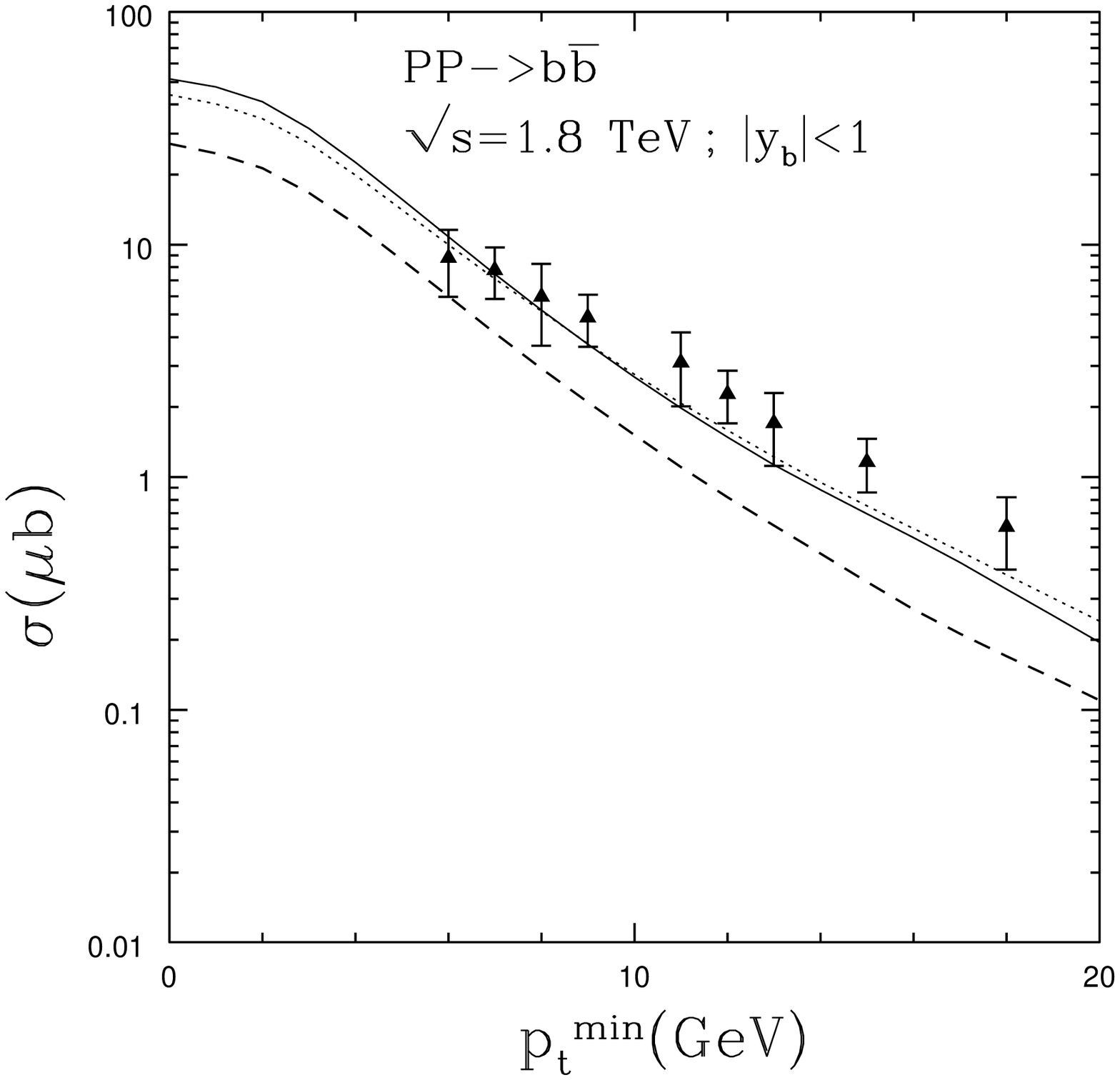,height=3.2in}
\epsfig{figure=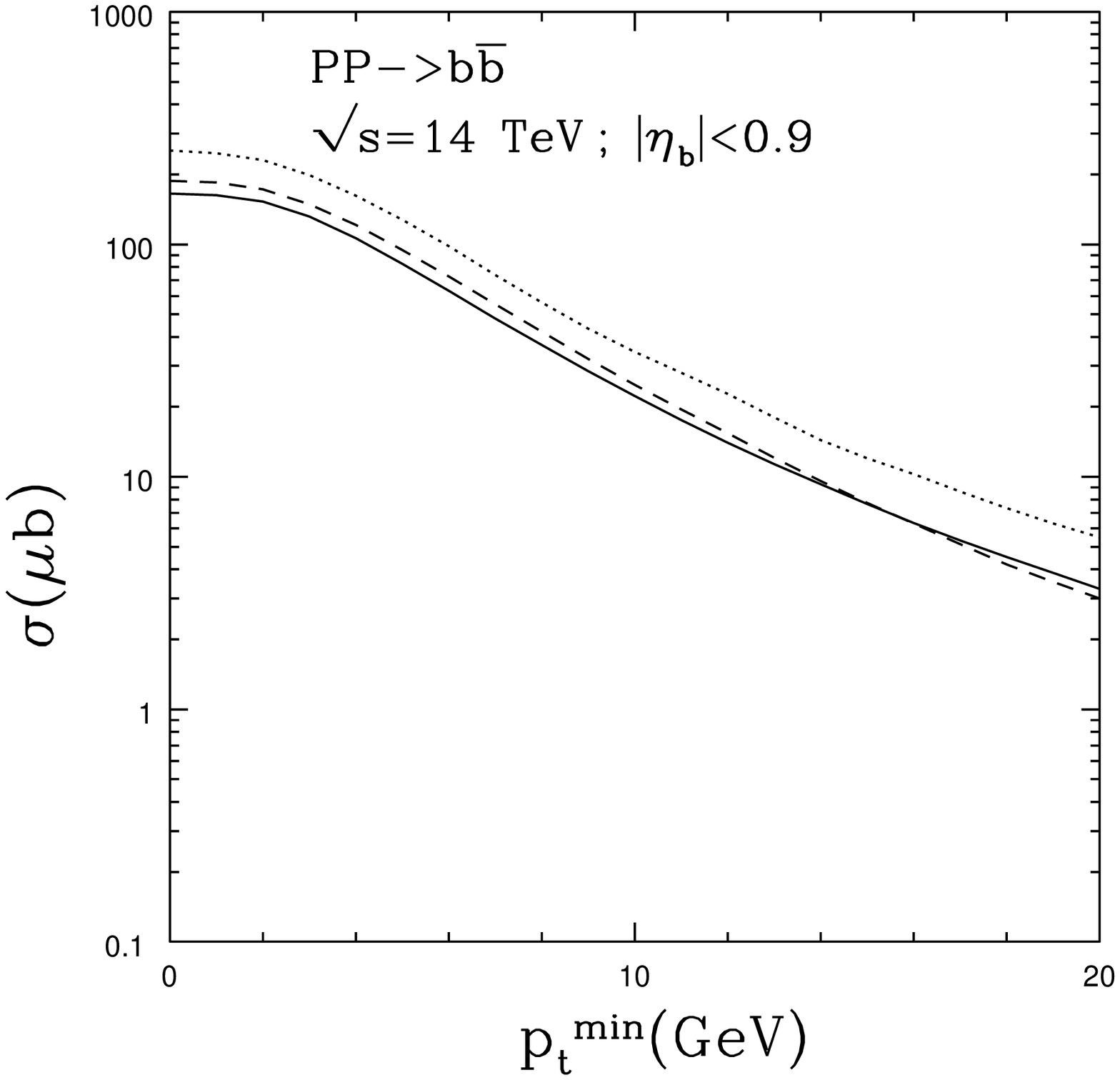,height=3.2in} \caption[ ]
{\footnotesize $p \bar p\rightarrow b\bar b$ production cross
section as a function of $p_t^{min}$ at  ${\sqrt s}=1.8$ TeV, with
the $b$-quark within the rapidity range $|y_b|< 1$, experimental
data from ref.\cite{Abbott:1999se}, and at ${\sqrt s}=14$ TeV with
the $b$-quark within the pseudo-rapidity range $|\eta|< 0.9$ .
\label{tev}  }
\end{center}
\end{figure}

In Fig.\ref{tev} we plot the integrated cross section of
 $b\bar b$ production at TEVATRON($\sqrt s=1.8$TeV) and at
LHC($\sqrt s=14$TeV), as a function of the minimum value of the
transverse momentum $p_t^{min}$ of the $b$-quark. The dotted
curves represent the cross section derived using the unintegrated
gluon structure function, according with the BFKL prescription of
Eq(4), whereas the dashed lines have been obtained by using the
prescription in Eq(3). The continuous lines represent the result
of the lowest order calculation multiplied by the $K$ factor. At
TEVATRON energy the $b$-quark distributions are within the
rapidity interval $|y|<1$ and are compared with the D0
experimental data ~\cite{Abbott:1999se}. The same distributions,
extrapolated at LHC energy, are then plotted as a function of
$p_t^{min}$ within the pseudorapidity interval $|\eta|<0.9$,
corresponding to the acceptance of the ALICE detector.

\begin{figure}[t]
\begin{center}
\epsfig{figure=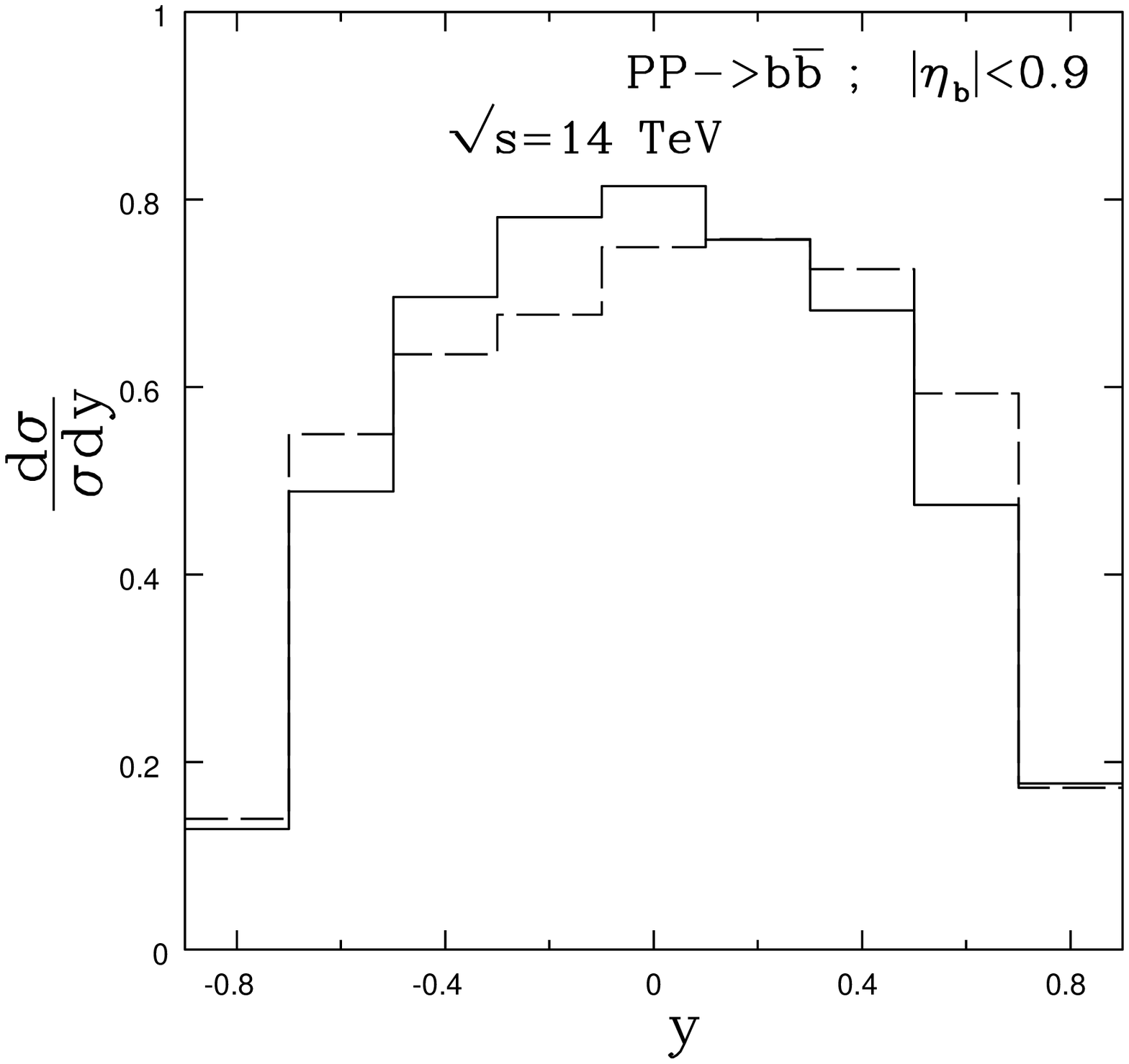,height=3.2in}
\epsfig{figure=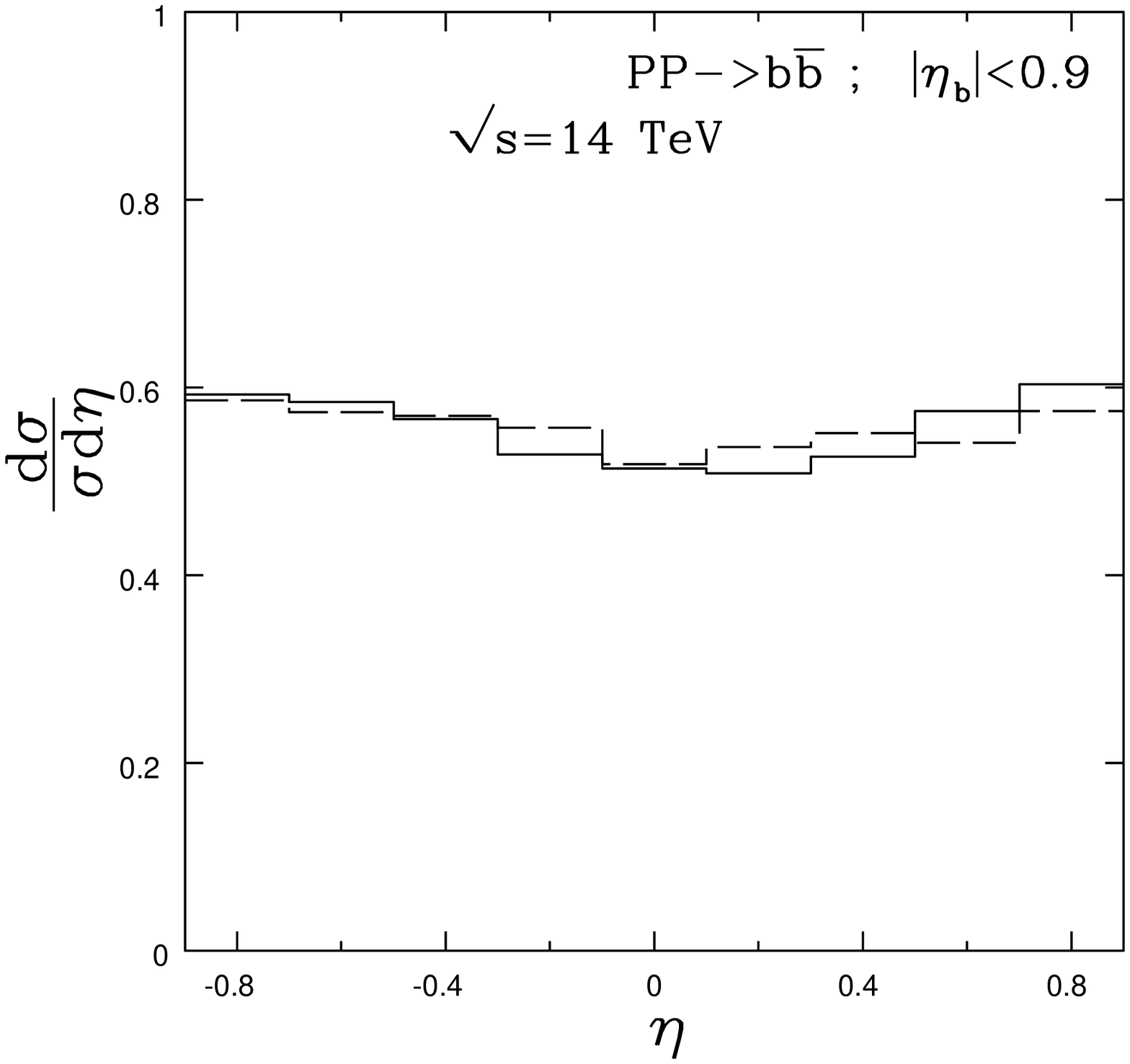,height=3.2in} \caption[ ]
{\footnotesize Normalized rapidity ($y$) and pseudorapidity
($\eta$) distributions for $b\bar b$ production at ALICE. With the
$k_t$ factorization approach (dashed histograms) and at the lowest
order in pQCD multiplied by the $K$-factor (continuous
histograms). \label{etapm} }
\end{center}
\end{figure}

In  Fig.\ref{etapm} we show the rapidity$(y)$ and
pseudorapidity$(\eta)$ distributions normalized to one and within
$|\eta|<0.9$. Here the continuous histograms refer to the result
of the lowest order calculation, rescaled by the $K$ factor,
whereas the dashed histograms represent the distributions
evaluated with the $k_t$-factorization approach. As one may see
also in this case the whole effect of higher orders reduces to a
simple rescaling.

\section{$ b{\bar b}b{\bar b}$ cross section }

The leading order QCD process to produce two pairs of heavy quarks
is given by the single parton scattering term at the fourth order
in the coupling constant\cite{Barger:1991vn}. A competing
mechanism at the LHC energy is the double parton
scattering\cite{DelFabbro:2000ds}. We compare the two mechanisms
in proton-proton collisions in the kinematical range of the ALICE
and of the LHCb detectors, namely at center-of-mass energies of
$5.5$ and $14$ TeV, within the pseudorapidity regions $|\eta|<0.9$
and $1.8<\eta<4.9$, down to very low transverse momenta.

The single scattering pQCD sub-processes at the lowest order in
$\alpha_s$, in $pp\rightarrow b\bar b b \bar b$, are the quarks
initiated process, $q \bar q\rightarrow b\bar b b \bar b$, whose
amplitude is given by the sum of $14$ Feynman diagrams for each
flavor in the initial state, and gluon fusion, $g g\rightarrow
b\bar b b \bar b$, represented by $76$ diagrams altogether, the
latter amplitude giving the dominant contribution to the cross
section at small $x$. To evaluate the cross section we have
generated the matrix elements of the partonic amplitudes with
MadGraph~\cite{madgraph} and HELAS~\cite{helas} and we have used
the MRS99 parton distributions~\cite{mrs99}, with the
factorization scale equal to the renormalization scale
$\mu_F=\mu_R$, which we have kept fixed at the value of the
transverse mass of the produced $b$ quark. For the mass of the
bottom quark we have used the value $m_b=4.6$ GeV. The
multi-dimensional integrations have been performed by
VEGAS~\cite{vegas} and the resulting cross section has been
finally multiplied by the $K$ factor obtained as described in the
previous section.

The evaluation of the double parton scattering contribution to the
cross section is considerably more uncertain because of the
unknown non perturbative input to the process, given by the
two-body parton distribution functions
$\Gamma(x_1,x_2,\beta)$\cite{dpth}, where $x_{1,2}$ are the
fractional momenta of the two partons belonging to the same hadron
and $\beta$ their distance in transverse space. Although not
explicitly written, the distributions depend also on the scale
factors characterizing each elementary interaction and on the
different kinds of partons involved. Given the large parton
population at low $x$, to proceed further we make the usual
simplifying assumption of neglecting correlations in fractional
momenta and we factorize the two-body parton distribution as
$\Gamma(x_1,x_2,\beta)=G(x_1) G(x_2) F(\beta)$, where $G(x)$ are
the usual one-body parton distributions and $F(\beta)$ is a
function normalized to 1 and representing the parton pair density
in transverse space. With these assumptions the cross section
acquires the simplified form\cite{Calucci:1999yz}
\begin{equation}
\sigma_D(b{\bar b};b{\bar
b})=\frac{1}{2}\sum_{ij}\Theta^{ij}\sigma_{i}(b{\bar
b})\sigma_{j}(b{\bar b}) \label{sdouble}
\end{equation}
where the indices $i,j$ label the different cases where each
$b{\bar b}$ pair is originated either by a $q{\bar q}$
annihilation, discriminating the cases of sea and valence, or by
two gluons and $\sigma_{i}(b{\bar b})$ represents the inclusive
cross sections for $b{\bar b}$ production in a hadronic collision,
with the index $i$ labelling a definite parton process. The weight
factors $\Theta^{ij}$ have dimension an inverse cross section and
result from integrating the product of the two-body parton
distributions in transverse space, while the factor $1/2$ is a
consequence of the symmetry of the expression for exchanging $i$
and $j$. The dependence of $\Theta^{ij}$ on the indices $i,j$
accounts for the possibility, for different pairs of partons in
the hadron structure, to be characterized by different values of
their relative average transverse distance~\cite{Calucci:1999yz}
\cite{DelFabbro:2000ds}. Notice that by measuring the double
parton collisions one has access to a new information on the
hadron structure, summarized in these weight factors, which cannot
be obtained in hard processes with a single parton interaction
only.

The experimental information on the double parton scatterings is
due to the four-jet production measurement in $pp$ collisions at
$\sqrt s=63$ GeV, performed by the AFS
Collaboration\cite{Akesson:1986iv}, and to the study of final
states with three minijets and one photon in $p{\bar p}$
collisions at $\sqrt s=1800$ GeV, due to CDF
\cite{Abe:1997bp}\cite{Abe:1997xk}. In both cases the cross
section was expressed as
\begin{equation}
\sigma_D=\frac{m}{2}\frac{\sigma_S(A)\sigma_S(B)}{\sigma_{eff}}
\label{cdf}
\end{equation}
where $m=1$ if the two parton processes $A$ and $B$ are identical,
while $m=2$ if they are different and $\sigma_S$ is the single
scattering inclusive cross section. The overall output of the
experiment hence reduces to the value of a single parameter, the
scale factor $\sigma_{eff}$, whose value is $\sigma_{eff}=5$ mb
for AFS, while $\sigma_{eff}=14.5$ mb for CDF. The two
experimental results are not inconsistent, given the different
content of partons in the two cases, mainly valence quarks in the
former case and mostly gluons and sea quarks in the latter; the
experimental indication hence pointing in the direction of a
sizable dependence of the factors $\Theta^{ij}$ in (\ref{sdouble})
on the different elementary processes. Interestingly, the
measurement of $\sigma_{eff}$ for different final states, as a
function of the c.m. energy and cuts applied, allows one obtaining
the values of the scale factors $\Theta^{ij}$, letting in this way
access to the three dimensional structure of the
proton\cite{DelFabbro:2000ds}.

The dominant contribution to $b{\bar b}b{\bar b}$ production at
the LHC is gluon fusion, so, for the present purposes, the sum in
(\ref{sdouble}) may be approximated well by a single term, where
the scale factor could not be strongly different with respect to
the case of the CDF experiment. Hence, to evaluate the double
scattering cross section, we have used the simplest expression

\begin{equation}
\sigma_D(b{\bar b}b{\bar b})=\frac{\sigma(b{\bar
b})^2}{2\sigma_{eff}} \,. \label{fact}
\end{equation}
where for $\sigma_{eff}$ we have taken the value reported by CDF.
Notice that since $\sigma_D$ is proportional to $\sigma_S^2$, the
effect of higher order corrections is enhanced on $\sigma_D$:
\\
\begin{eqnarray}
\sigma_S\,=\,K\,\sigma_S^{LO}\nonumber\\
\sigma_D\,=\,K^2\,\sigma_D^{LO}
\label{ksig}
\end{eqnarray}
where $\sigma_{S,D}^{LO}$ refers to the lowest order expressions
of the cross section.

\section{ Results}

\begin{figure}
\vspace{.1cm} \centerline{ \epsfysize=9cm \epsfbox{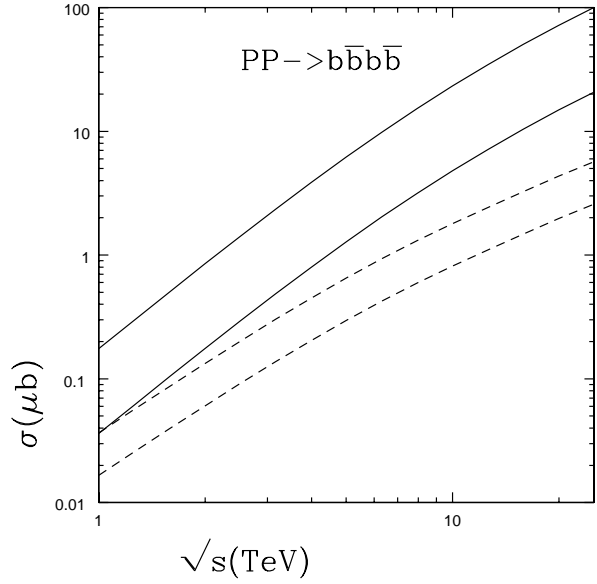}}
\vspace{.1cm} \caption[ ]{ $b\bar b b \bar b$ total cross section
as a function of centre of mass energy. Lower curves $K=2.5$,
higher curves $K=5.5$ .} \label{4becm}
\end{figure}

In Fig.\ref{4becm} we plot the expected rise of the total $b{\bar
b}b{\bar b}$ production cross section as a function of the c.m.
energy. The continuous curves refer to the double parton
scattering contribution, while the dotted curves to single
scattering. In each case the lower curve refers to the value
$K=2.5$, while the higher curve to $K=5.5$, which are the typical
estimate of the NLO-QCD and the result of our calculation within
the $k_t$-factorization approach. Notice that at the LHC the
double parton scattering gives a contribution to the integrated
cross section about ten times larger than the single scattering.

\begin{figure}[t]
\begin{center}
\epsfig{figure=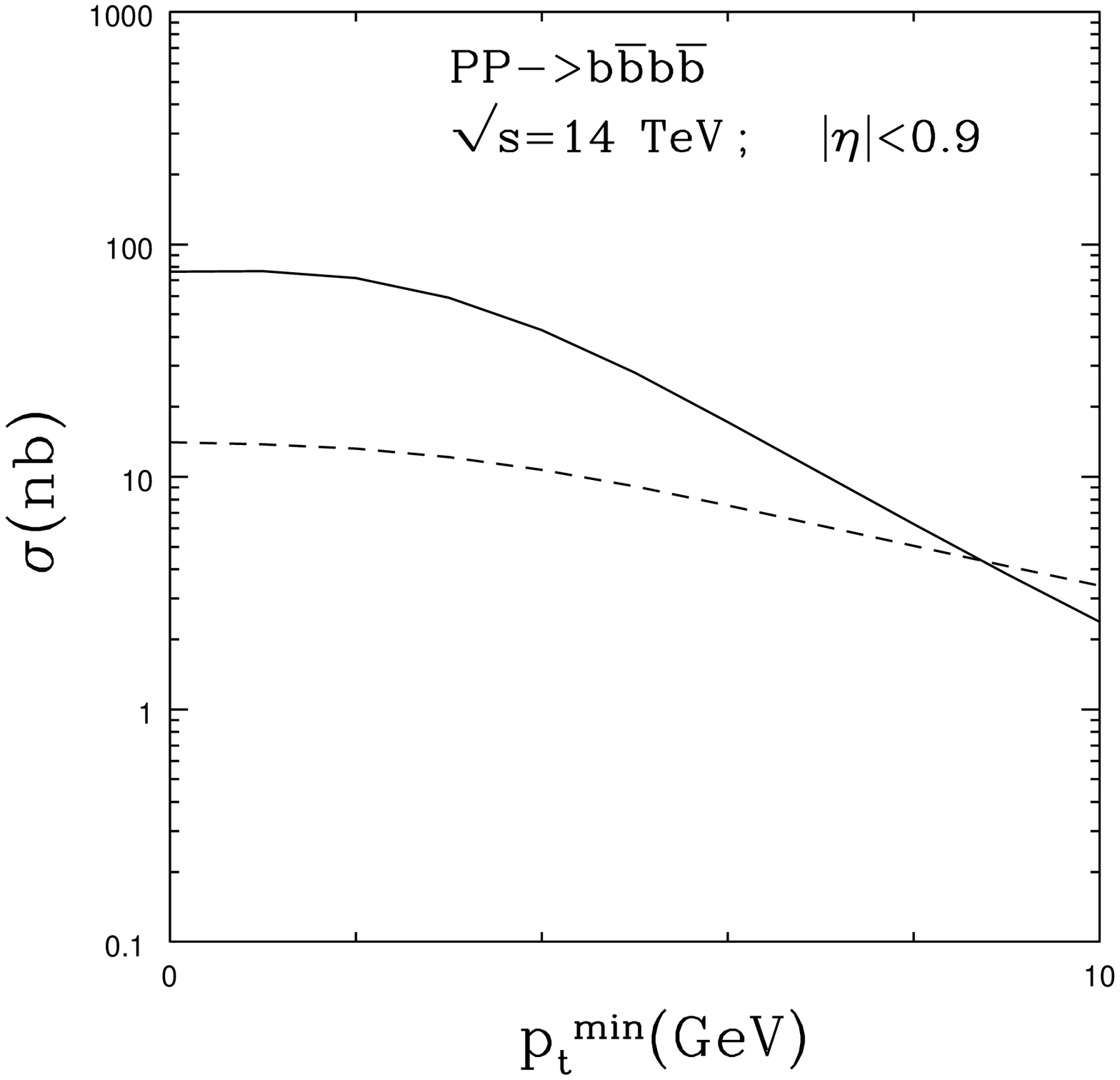,height=3.2in}
\epsfig{figure=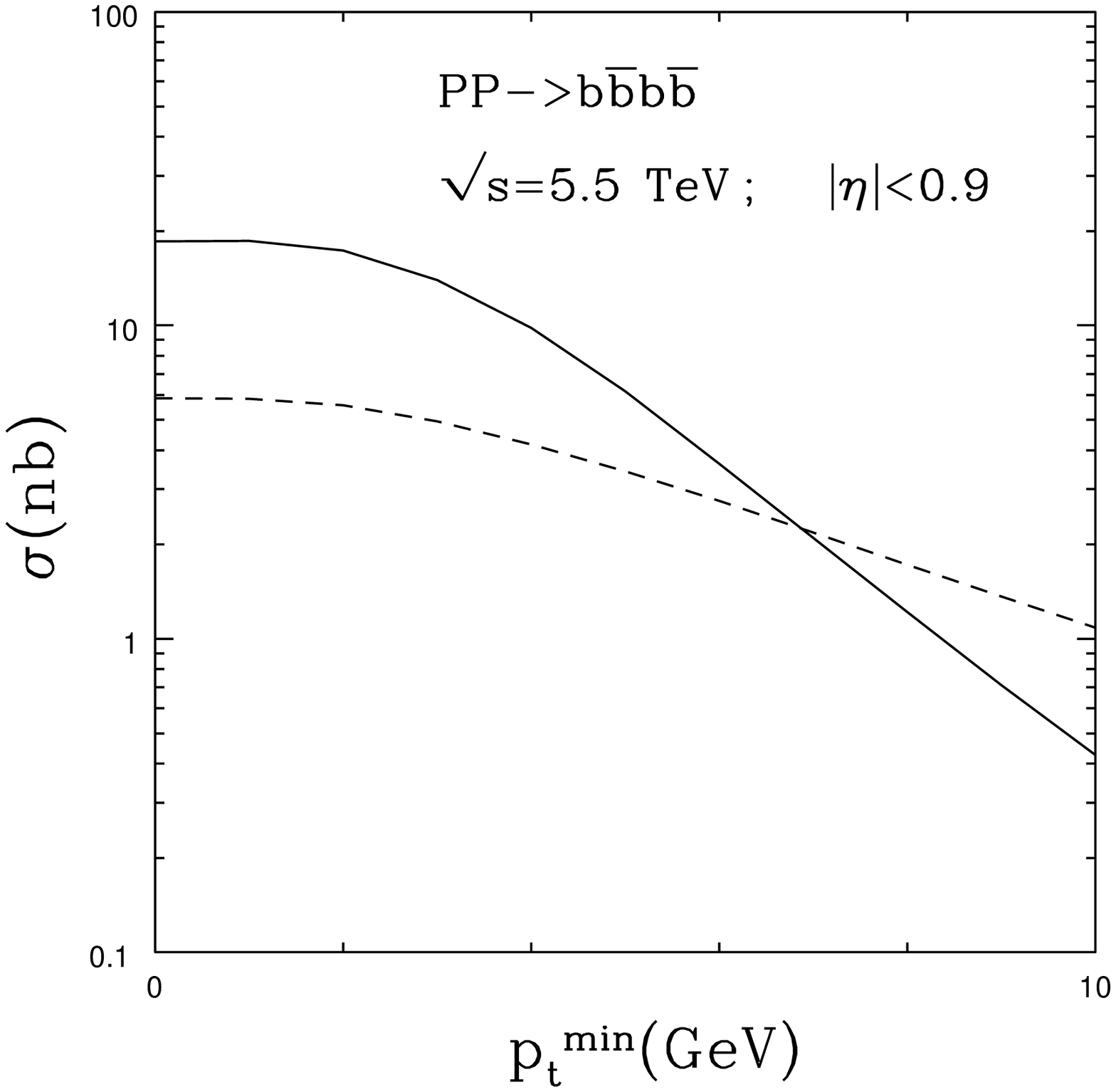,height=3.2in} \caption[ ]
{\footnotesize $b\bar b b \bar b$ production cross section at
${\sqrt s}=14$ TeV  and  at ${\sqrt s}=5.5$ TeV  as a function of
$p_t^{min}$ with all the four $b$-quarks in the pseudo-rapidity
interval $|\eta|< 0.9$ .} \label{4bpt14}
\end{center}
\end{figure}

\begin{figure}[t]
\begin{center}
\epsfig{figure=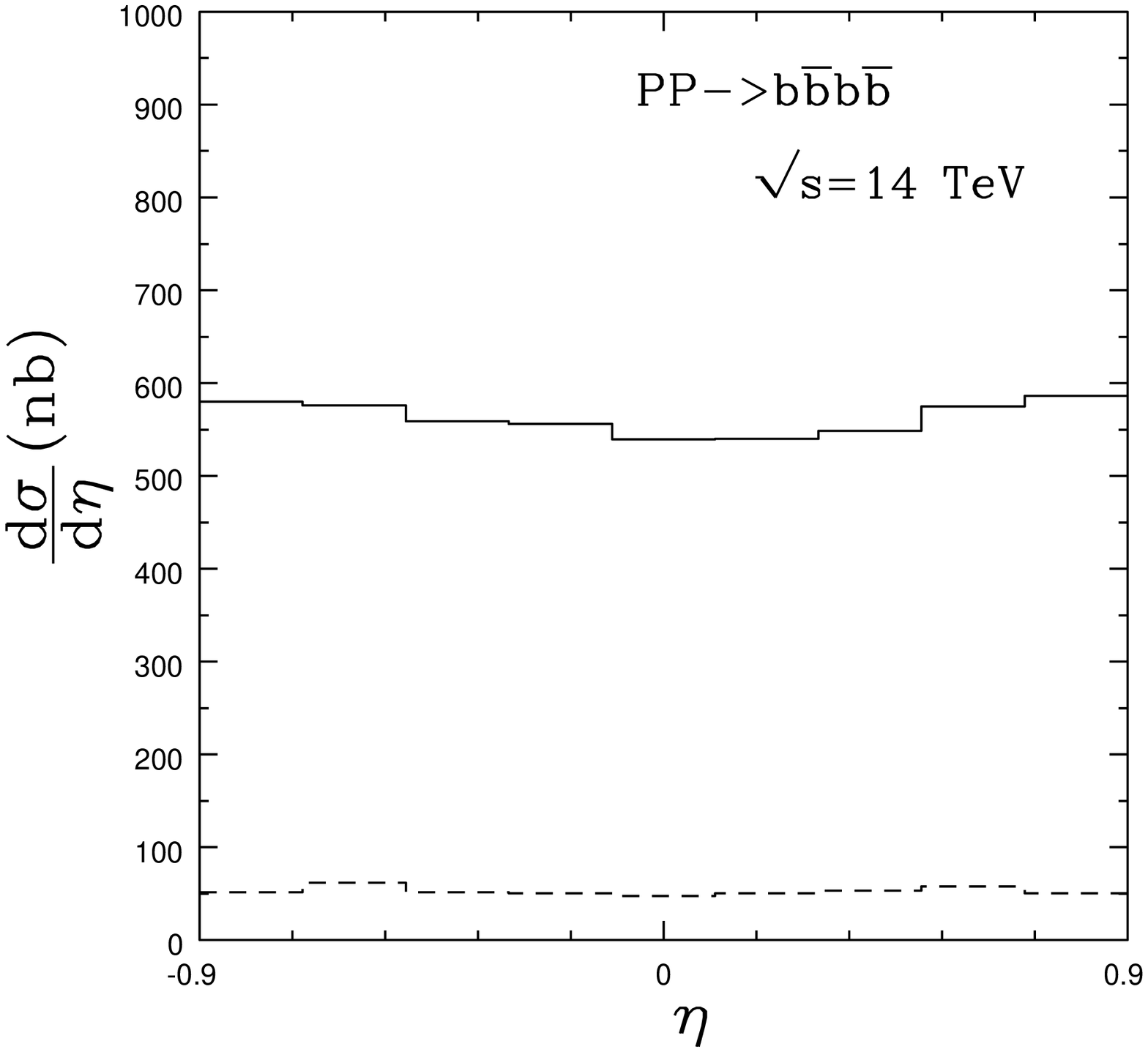,height=3.2in}
\epsfig{figure=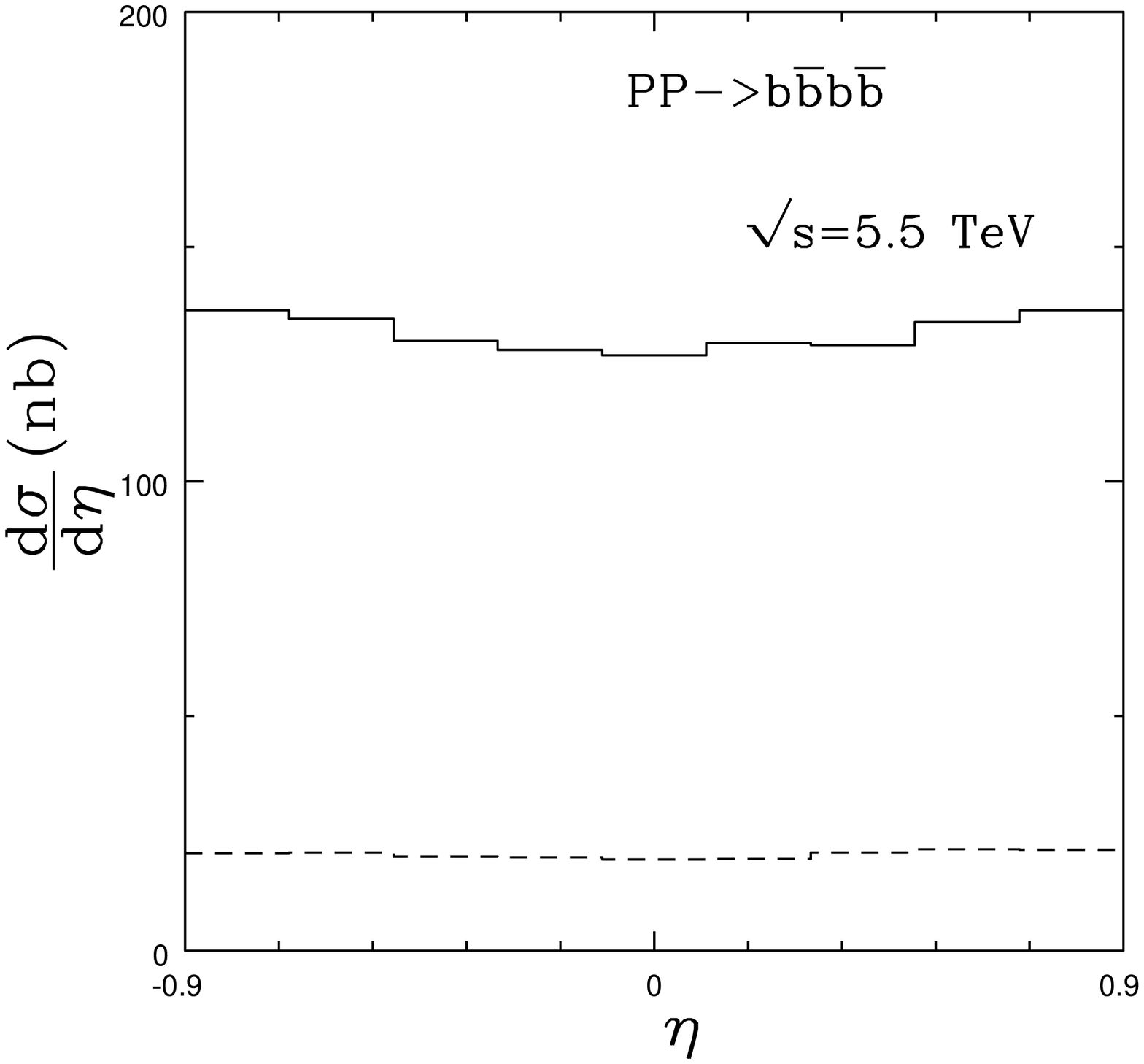,height=3.2in}
\epsfig{figure=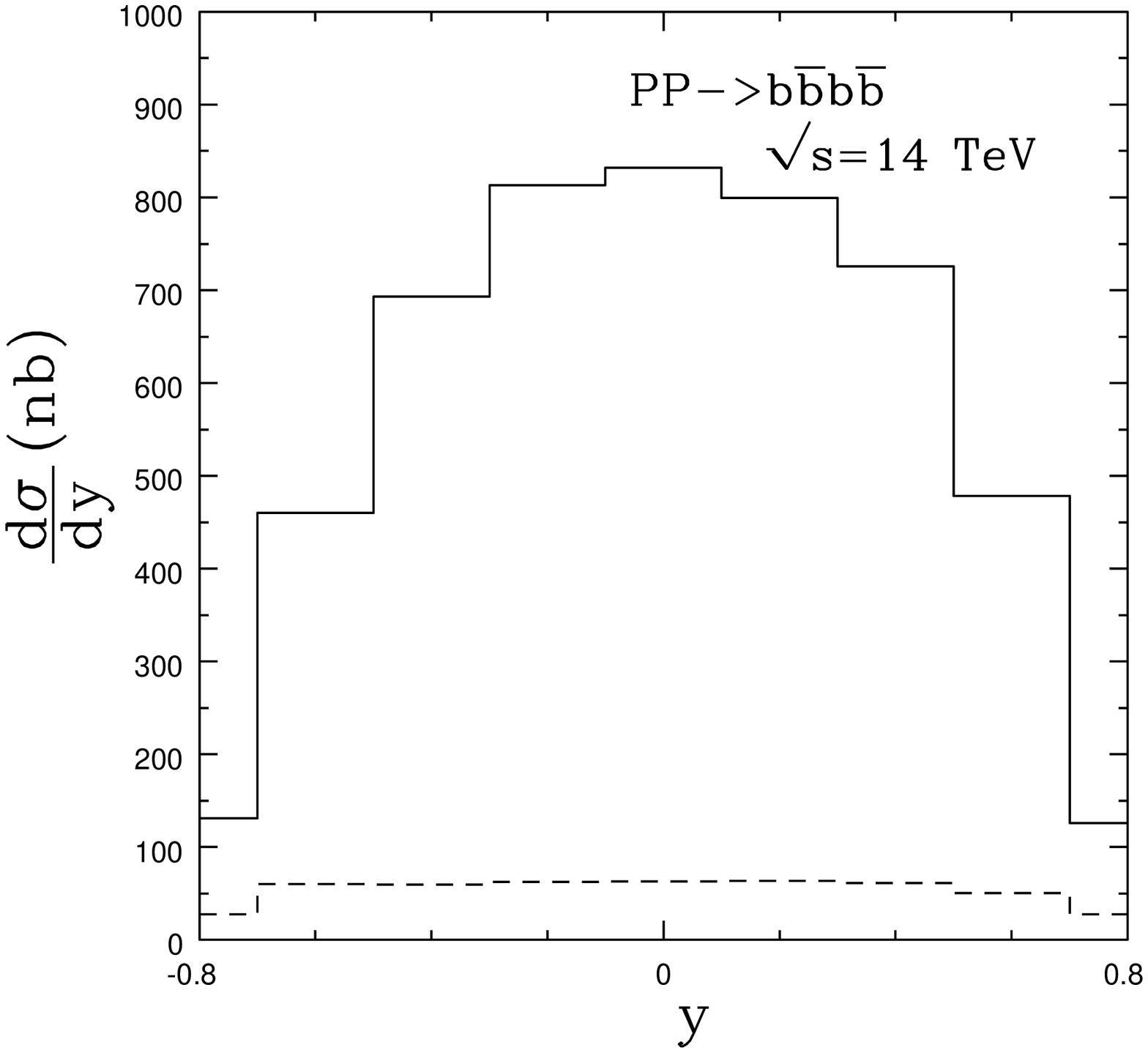,height=3.2in}
\epsfig{figure=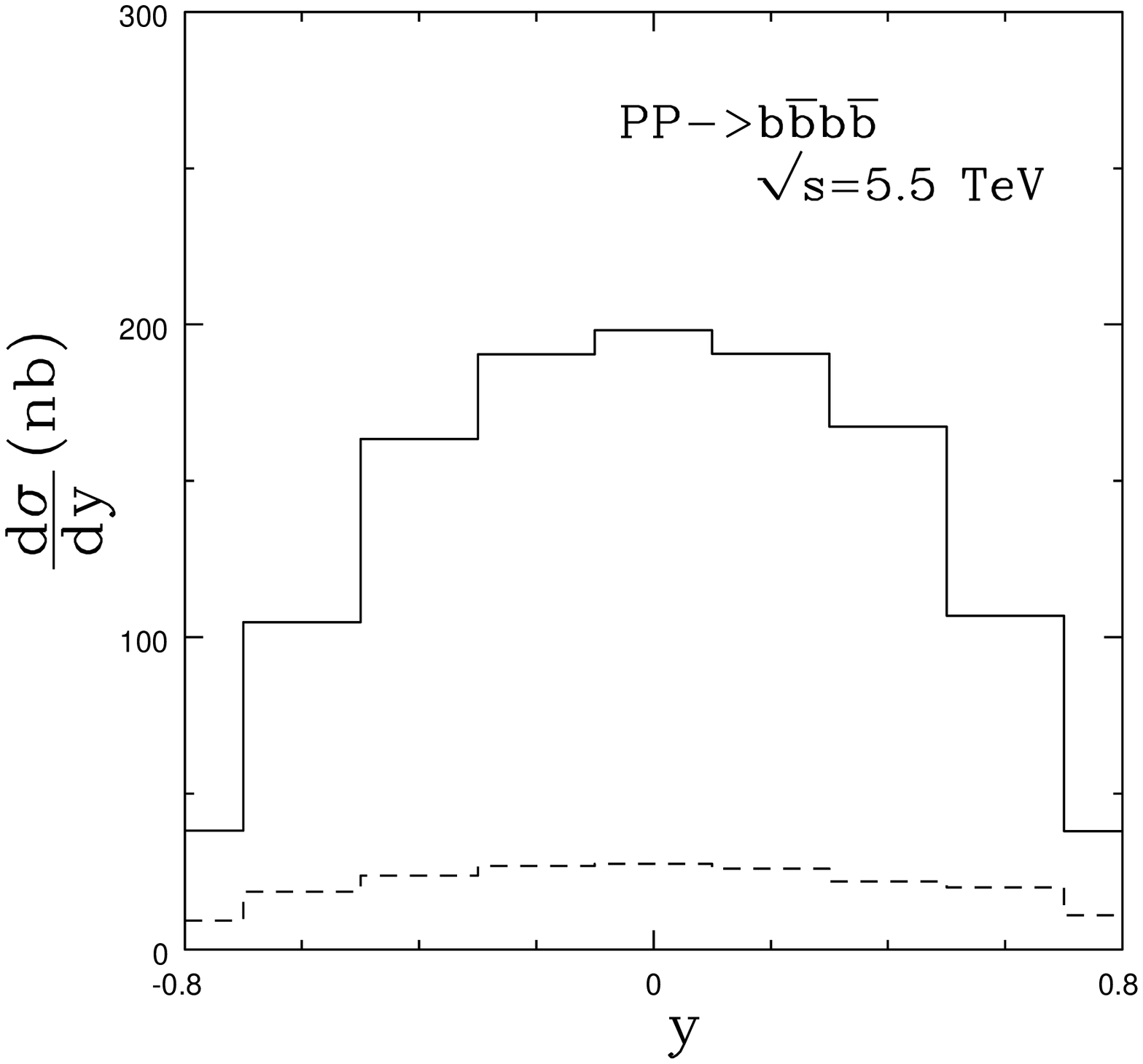,height=3.2in} \caption[ ]
{\footnotesize $b\bar b b \bar b$  production with the two equal
sign $b$-quarks in the pseudo-rapidity interval $|\eta_b|<0.9$.
$\eta$-distributions and $y_b$-distributions  at ${\sqrt s}=14$
TeV and at ${\sqrt s}=5.5$ TeV. The continuous histograms refer to
the contribution of double parton scatterings while the dashed
histograms to the single parton scatterings.} \label{4beta14}
\end{center}
\end{figure}

To see how the cross section depends on the transverse momenta we
have plotted in Fig.\ref{4bpt14} the two contributions to the
integrated cross section, as a function of $p_t^{min}$, the
minimum value of the transverse momenta of the $b$ quarks (which
we require to be all inside the pseudorapidity interval
$|\eta|<0.9$), for the center-of-mass energy  values of 14
and 5.5 TeV. The continuous histograms refer to the double parton
scattering contribution, while the dotted histograms to single
scattering. The double parton cross section decreases faster with
$p_t^{min}$ than the single parton cross section, the two
contributions being of the same order at $p_t^{min}=8-10$ GeV.
Pseudorapidity, and rapidity distributions at 14 and 5.5 TeV are
plotted in Fig.\ref{4beta14} ( always requiring for the two
$b$-quarks $|\eta|<0.9$), where continuous and dashed histograms
have the same meaning as in the previous cases.

\begin{figure}[t]
\begin{center}
\epsfig{figure=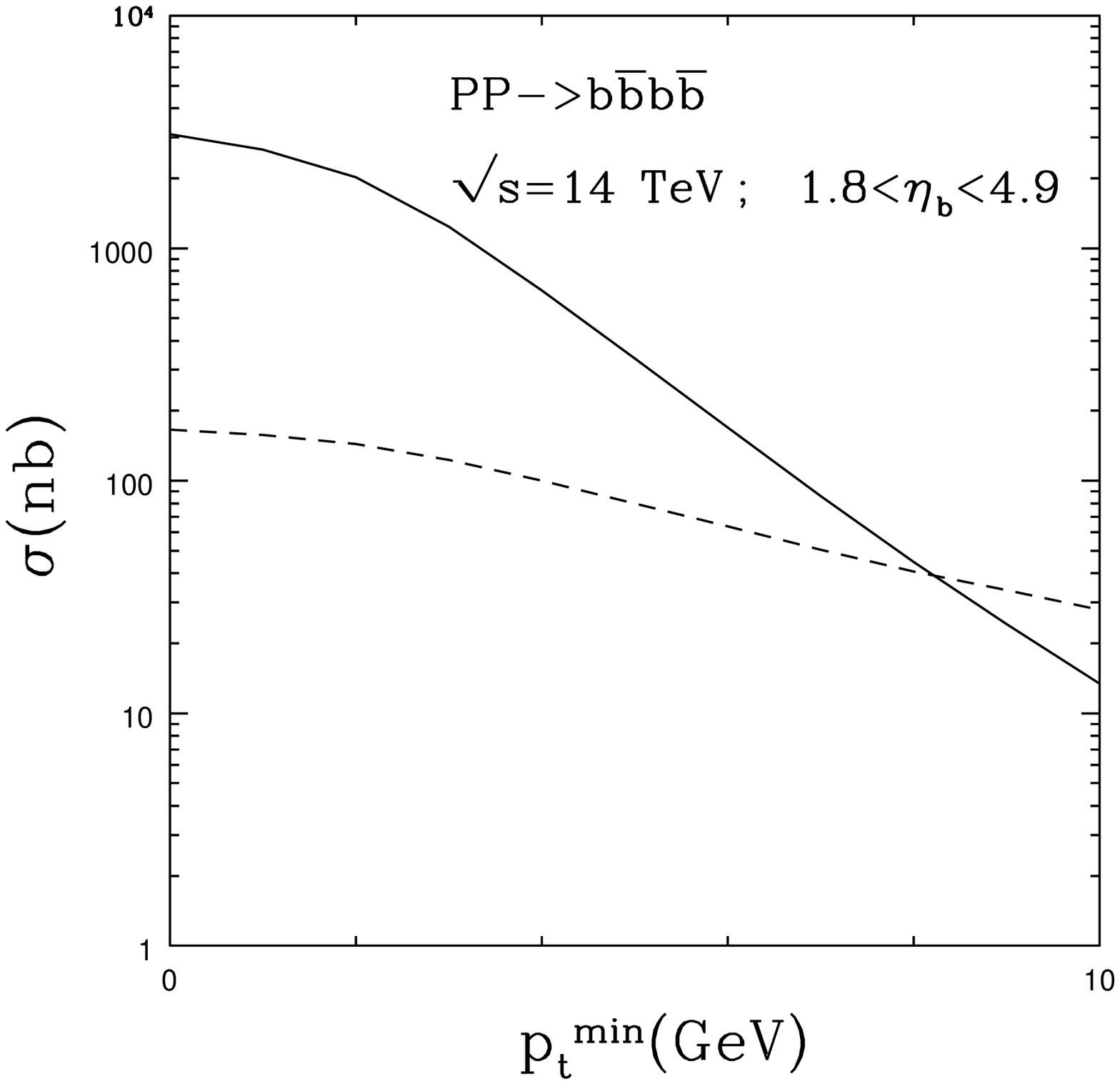,height=3.2in}
\epsfig{figure=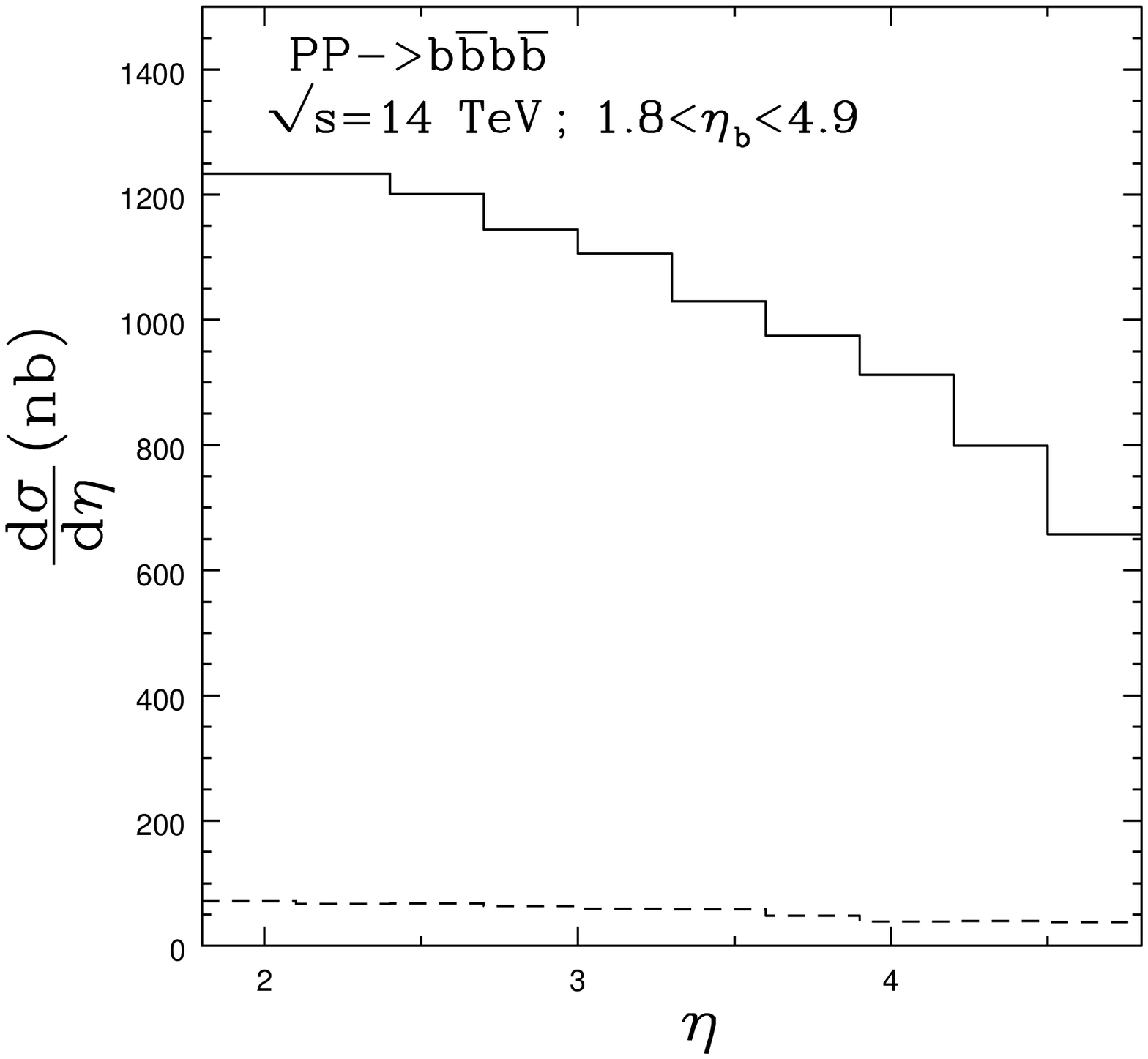,height=3.2in}
\epsfig{figure=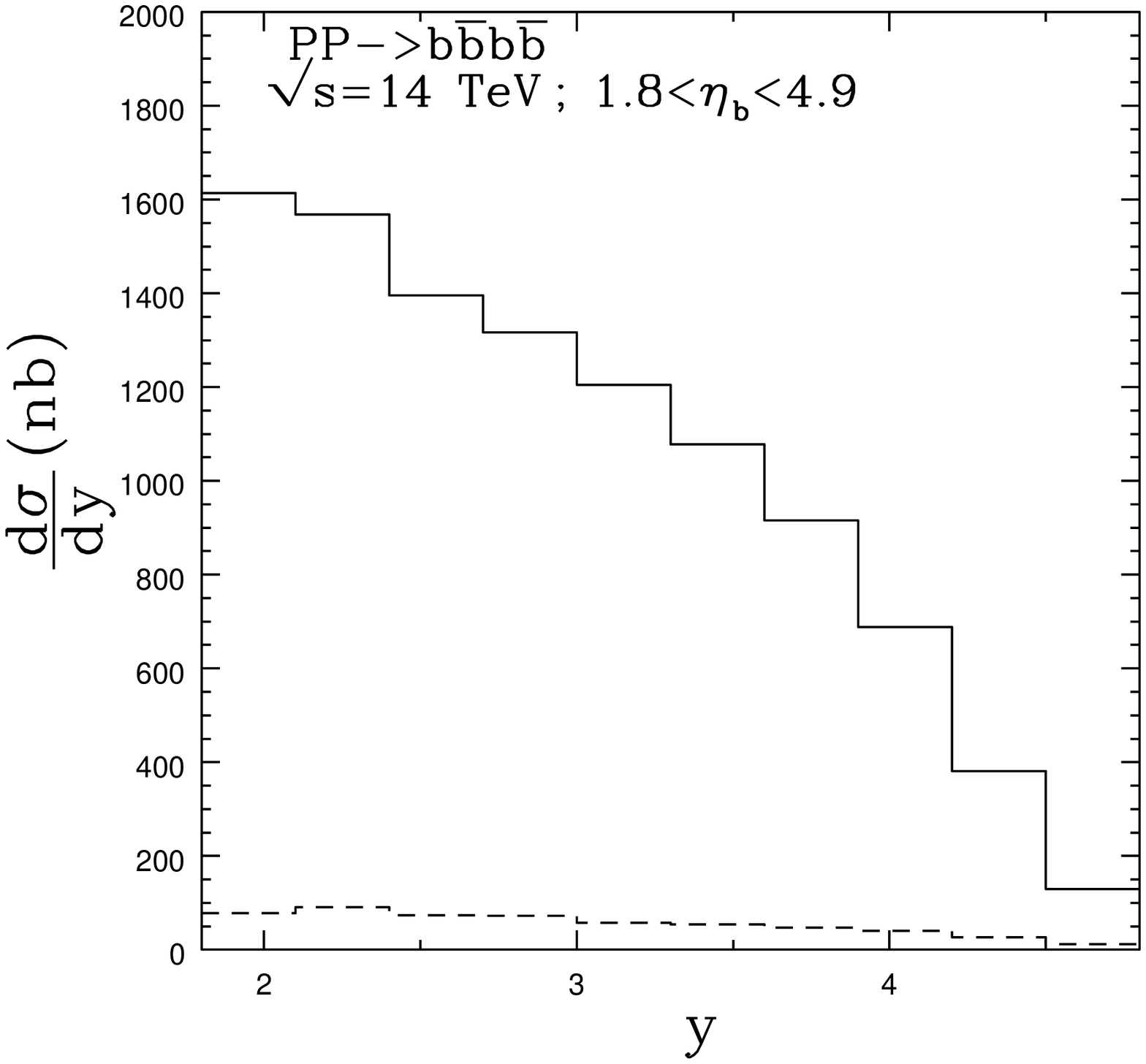,height=3.2in} \caption[ ]
{\footnotesize $b\bar b b \bar b$  production with the two equal
sign $b$-quarks in the pseudo-rapidity interval $1.8<\eta<4.9$
  at ${\sqrt s}=14$ TeV.
Production cross section as a function of $p_t^{min}$, $\eta$
 and $y$. The continuous lines and histograms refer to
the contribution of double parton scatterings while the dashed
lines and histograms to the single parton scatterings.} \label{4blhcb}
\end{center}
\end{figure}

To see how the results depend on rapidity, we have plotted in
Fig\ref{4blhcb} the same rapidity ($y$) and pseudorapidity
($\eta$) distributions at $\sqrt s =14$ TeV, requiring both
$b$-quarks to be in the pseudorapidity interval $1.8<\eta<4.9$,
which corresponds to the acceptance of the LHCb experiment. In the
same figure we also compare the two contributions of single and
double parton scattering, integrated within the rapidity
acceptance of the LHCb, as a function of $p_t^{min}$.

The overall indication which one obtains from the present study is
that double parton scatterings dominate the $b{\bar b}b{\bar b}$
integrated cross section by a large factor, both in the central
rapidity region and at the larger rapidity values of the LHCb
experiment. In both cases the contribution of the single parton
scattering term becomes important only after applying cuts to the
transverse momenta of the order of 8-10 GeV. The large values
expected for the $b{\bar b}b{\bar b}$ cross section, which at $14$
TeV are of the order of one $\mu$b, inside the ALICE and the LHCb
detectors, and the localization of the double scattering
contribution at relatively low $p_t$ values, suggest that heavy
quark pairs production at the LHC might represent an efficient
tool for studying the gluon initiated double parton scattering
process.

\vskip.25in {\bf Acknowledgment} \vskip.15in This work was
partially supported by the Italian Ministry of University and of
Scientific and Technological Researches (MIUR) by the Grant
COFIN2001.

\end{document}